\documentclass[aps, prapplied]{revtex4-2}

 \pdfoutput=1
\usepackage{amsfonts}
\usepackage{amsmath}
\usepackage{amssymb}
\usepackage{amstext}
\usepackage{amsthm}
\usepackage{appendix}
\usepackage{bm}
\usepackage[justification=raggedright,singlelinecheck=false]{caption}
\usepackage{color}
\usepackage{dcolumn}
\usepackage{empheq}
\usepackage{float}
\usepackage{gensymb}
\usepackage{graphicx}
\usepackage{mathtools}
\usepackage{mathrsfs}
\usepackage{multirow} 
\usepackage[final]{showkeys}
\usepackage{subfigure}
\usepackage{textcomp}
\usepackage{textcase}
\usepackage{units}
\usepackage{xr}
\usepackage[]{inputenc}

\usepackage{sanitize-umlaut}
\usepackage[switch,columnwise]{lineno}
\usepackage{datetime}
\newdate{date}{23}{02}{2024} 

\newcommand{\werb}[3]{\unit[(#1 $\pm$ #2)]{#3}}  

\newcommand{\abs}[1]{\left\vert #1 \right\vert}

\newcommand{\NCNR}{NIST Center for Neutron Research (NCNR)\renewcommand{\NCNR}{NCNR}}
\newcommand{\NIST}{National Institute of Standards and Technology (NIST)\renewcommand{\NIST}{NIST}}
\newcommand{\NIOF}{Neutron Interferometry and Optics Facilities (NIOF)\renewcommand{\NIOF}{NIOF}}
\newcommand{\MGH}[1]

\newcommand\filledcirc{\ensuremath{{\color{blue}\bullet}\mathllap{\color{blue}\circ}}}

\begin{document}

\title{Achieving a Near-Ideal Silicon Crystal Neutron Interferometer using Sub-Micron Fabrication Techniques}

\author{M. G. Huber} 
\email{michael.huber@nist.gov}
\affiliation{Physical Measurement Laboratory, National Institute of Standards and Technology, Gaithersburg, MD 20899}
\author{I. Taminiau} 
\affiliation{Neutron Optics Inc.,Waterloo, ON M5N 2N1}
\author{D. G. Cory}
\affiliation{Department of Chemistry, University of Waterloo, Waterloo, ON N2L 3G1} 
\affiliation{Institute for Quantum Computing, University of Waterloo, Waterloo, ON N2L 3G1}
\author{B. Heacock}
\affiliation{Physical Measurement Laboratory, National Institute of Standards and Technology, Gaithersburg, MD 20899}
\author{D. Sarenac}
\affiliation{Department of Physics, University of Waterloo, Waterloo, ON N2L 3G1} 
\affiliation{Department of Physics, University at Buffalo, Buffalo, NY 14260}
\author{R. Valdillez}
\affiliation{Department of Physics, North Carolina State University, Raleigh, NC 27695} 
\affiliation{Triangle Universities Nuclear Laboratory, Durham, NC 27708}

\author{D. A. Pushin}
\affiliation{Department of Physics \& Astronomy, University of Waterloo, Waterloo, ON N2L 3G1}
\affiliation{Institute for Quantum Computing, University of Waterloo, Waterloo, ON N2L 3G1}


\begin{abstract}
Perfect-crystal neutron interferometry which is analogous to Mach-Zehnder interferometry, uses Bragg diffraction to form interfering neutron paths. 
The measured phase shifts can be used to probe many types of interactions whether it be nuclear, electromagnetic, gravitational, or topological in nature.   
For a perfect-crystal interferometer to preserve coherence, the crystal must possess a high degree of dimensional tolerance as well as being relatively defect-free with minimal internal stresses. 
In the past, perfect-crystal neutron interferometers have been produced by a two step process. First, a resin diamond wheel would be used to remove excess material and shape the interferometer.  Afterword, the crystal would be etched in order to remove surface defects and elevate strains.
This process has had limitations in terms of repeatability and in maximizing the final contrast, or fringe visibility, of the interferometer. 
We have tested various fabrication and post-fabrication techniques on a single perfect-crystal neutron interferometer and measured the interferometer's performance at each step.
Here we report a robust, repeatable non-etching fabrication process with high final contrast.    
For the interferometer used in this work, we achieved contrasts of greater than \unit[90]{\%} several separate times and ultimately finished with an interferometer that has  \unit[92]{\%} contrast and a uniform phase distribution.
\end{abstract}
\keywords{perfect crystal neutron interferometry; advanced fabrication; silicon crystal}
\date{\displaydate{date}}

\maketitle{}

 \begin{figure}[t]
 \centering
\includegraphics[width = \columnwidth]{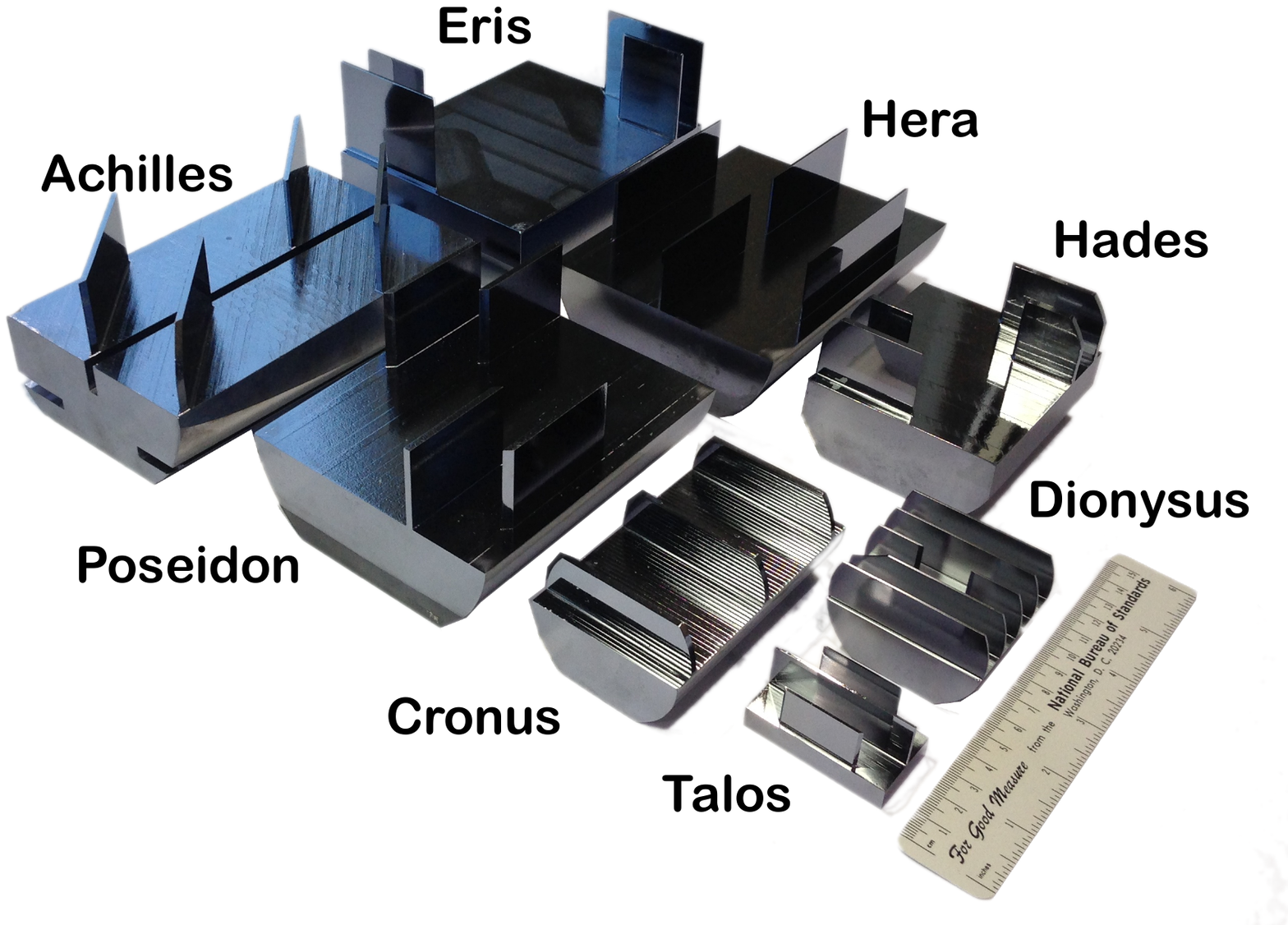} 
 \caption{Interferometer crystals. Hera, the largest LLL crystal, is  approximately 10 cm $\times$ 10 cm in size.  Talos, the subject of this work, is the smallest. }
 \label{inters}
\end{figure}

\begin{figure}[t]
 \centering
\includegraphics[width = \columnwidth]{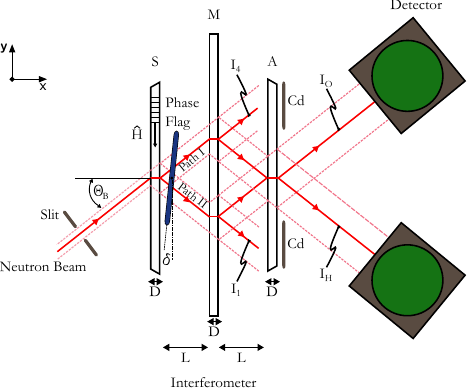} 
 \caption{Neutrons entering the interferometer are split via the first diffracting blade (S) into 2 paths.  The paths are again diffracted by a mirror blade (M).  Lastly,  the 2 paths interfere in an analyzer crystal (A).  Any differences between the paths modulates the intensities $I_O$ and $I_H$ measured at the detectors. By rotating a phase flag by an angle of $\delta$ one can  vary $I_O$ and $I_H$. Normally,  the distance $L$ is significantly long enough such that the $I_1$ and $I_4$ beams escape the interferometer without striking the analyzer crystal.  This is not the case for the interferometer Talos.  Here  neutron absorbing cadmium (Cd) sheets placed behind the analyzer crystal to eliminate $I_1$ and $I_4$ from reaching the detectors.}
 \label{BabyScheme}
\end{figure}

\section{Introduction}
Neutron interferometry utilizes the wave-like properties of neutron beams to study aspects of fundamental and applied physics. 
It has been used to demonstrate $4\pi$ spinor rotation \cite{Rauch1975}, orbital angular momentum \cite{Clark2015,Sarenac_2016_Opt.Express},
berry phases \cite{Werner2012}, Cheshire cats and weak measurements \cite{Danner2023},
GHZ states \cite{Erd_2013_NewJournalofPhysics}, probe nuclear structure \cite{Haun2020}, and search for beyond the Standard Model physics \cite{Heacock2021,Lemmel2015,Li2016}.
The neutron's unique sensitivity is due to its charge neutrality, magnetic dipole moment and non-zero spin, and its inherent quantum nature.
Measuring the neutron's phase allows precision and sensitivity not found in other neutron scattering techniques.
\par
Perfect-crystal neutron interferometers were first produced in the mid-1970s \cite{Rauch1974, Colella1975} following the introduction of x-ray interferometers \cite{Bonse1965} a decade before. 
Until recently the techniques for fabricating interferometer crystals has had little innovation \cite{Treimer1979, Zawisky2009}. 
In order to minimize impurities, interferometers are constructed from float-zone (FZ) grown \cite{ZULEHNER20007,MUIZNIEKS2015241} silicon ingots. 
The ingots are machined such that there are multiple crystal blades extruding from a common base.  
The large common base is necessary for stability and, most importantly, to align each crystal blade relative to each other at the sub-arc second level. 
Only very recently has a neutron interferometer been shown to operate with blades on separate bases \cite{Lemmel2022}. 
A collection of neutron interferometers are shown in Fig.~\ref{inters}.
\par
Traditionally, the ingots are machined into the correct shape using a resin diamond wheel (200 grit to 600 grit) \cite{ALLMAN1998392} with an accuracy of 10's of microns. 
The machining damages the crystal blade surface and creates microcracking that can penetrate deep into the subsurface \cite{Zhong2003,Smith_1990,YAN2009378}.  
Microcracking, subsurface stresses, and  other defects are removed by etching the crystal surfaces.
The etching process is done iteratively and removes only several microns of surface material at a time.
This is because excessive etching can decrease blade parallelism and has been shown to negatively impact the crystal's performance.
\par
For example, the crystal known  as Hades (see Fig.~\ref{inters}) was made in 1988 \cite{ClothThesis} at the University of Missouri-Columbia.  It was etched in a bath of hydrofluoric, acetic, and nitric acids at concentrations of \unit[18]{\%}, \unit[25]{\%}, and \unit[57]{\%} respectively.  
This process was repeated five times and a total of \unit[40]{$\mu$m} of material was removed from each blade (\unit[20]{$\mu$m} per side).
After each etch the contrast of Hades was measured.
The final maximum of contrast of Hades at that time being \unit[70]{\%}. 
\par
The etching of an interferometer has poor repeatability.  
The exothermic reactions taking place cause currents in the acid bath and results in uneven flow in the channels between the crystal blades.  
A greater issue is the formation of bubbles that adhere to the silicon surfaces. 
These issues cause variable etch rates along the interferometer blades that negatively effect the parallelism and geometry of the resulting neutron interferometer.  
The varying etch rate can be mediated, in part, by constantly moving the interferometer inside the bath.
Also, the acid bath can be `kick-started' by the introduction of several grams of silicon powder minutes before submerging the interferometer.
This starts the exothermic reaction and raises the bath temperature such that the etch rate is more constant in time.  
In practice, neither moving the crystal nor `kick-starting' the reaction can totally remove the factors that led to poor etching consistency. 
\par
Neutron interferometers make use of several different sizes and configurations. 
The construction of new interferometers is needed for next generation neutron interferometry experiments. 
Due to the expense of FZ material, the iterative machining process, and inconstancy of etching, a more dependable  fabrication process is needed.
Here we discuss sub-micron fabrication technique that can minimize and even eliminate the need for etching and yet still produce a high-contrast, high-quality interferometer.
These techniques were tested on an existing neutron interferometer.

\section{Neutron Interferometry}

In perfect-crystal neutron interferometry a neutron wave is manipulated using Bragg diffraction inside a crystal lattice. 
Figure \ref{BabyScheme} shows a schematic of a Laue-Laue-Laue (LLL) type interferometer. 
A mono-energetic neutron beam enters the interferometer and diffracts from the first blade (Splitter : S) resulting in the incident neutron beam splitting into two distinct paths: a forward diffracted (path I) and a reflected one (path II). 
Internal crystal blade(s) (Mirror : M) reflect the neutron paths so that they interfere with each other in a final blade (Analyzer : A).
This setup is analogous to a Mach-Zehnder interferometer in light optics. 
In reality, there is only one neutron entering the interferometer at time and it is the neutron's wave function that is shared between the two paths.
Or in other words, a neutron interferometer is a perfect example of self-interference. 
The exiting neutron beams labeled $I_O$ and $I_H$ for the O-beam and H-beam, respectively, are detected with near \unit[100]{\%} efficiency using $^3$He tubes. 
Differences in the phase that the neutron accumulates between paths I and II modulate the intensity at the detectors.
Phase shifts can be due to physical materials placed in the beam paths as well as forces and the particulars of the neutron's trajectory.
A device called a phase flag made from optically flat \unit[1.5]{mm} thick quartz is placed inside the interferometer.
Rotating the phase flag varies its effective thickness along each path resulting in a modulation of the outgoing $I_{O}$ and $I_{H}$ intensities.
\par
A selection of interferometer crystals of various sizes and geometries are shown in Fig.~\ref{inters}. 
Of the interferometers shown in Fig.~\ref{inters} only Hera had a maximum contrast above \unit[80]{\%} prior to this work.
As one can see in Fig.~\ref{inters}, a variety of crystal shapes and sizes are needed to meet the  requirements of a diverse set of experiments.
Recently, ultra precision machining has been utilized to cut a close to ideal 2-blade interferometer crystal \cite{Heacock2019}.
In addition, post-fabrication annealing has shown to alleviate internal crystal stresses and improve interferometer contrast \cite{Heacock2018}.

\subsection{Contrast}\label{Contrast}

\begin{figure}[t]
 \centering
\includegraphics[trim={0 0 0 0},clip, width = \columnwidth]{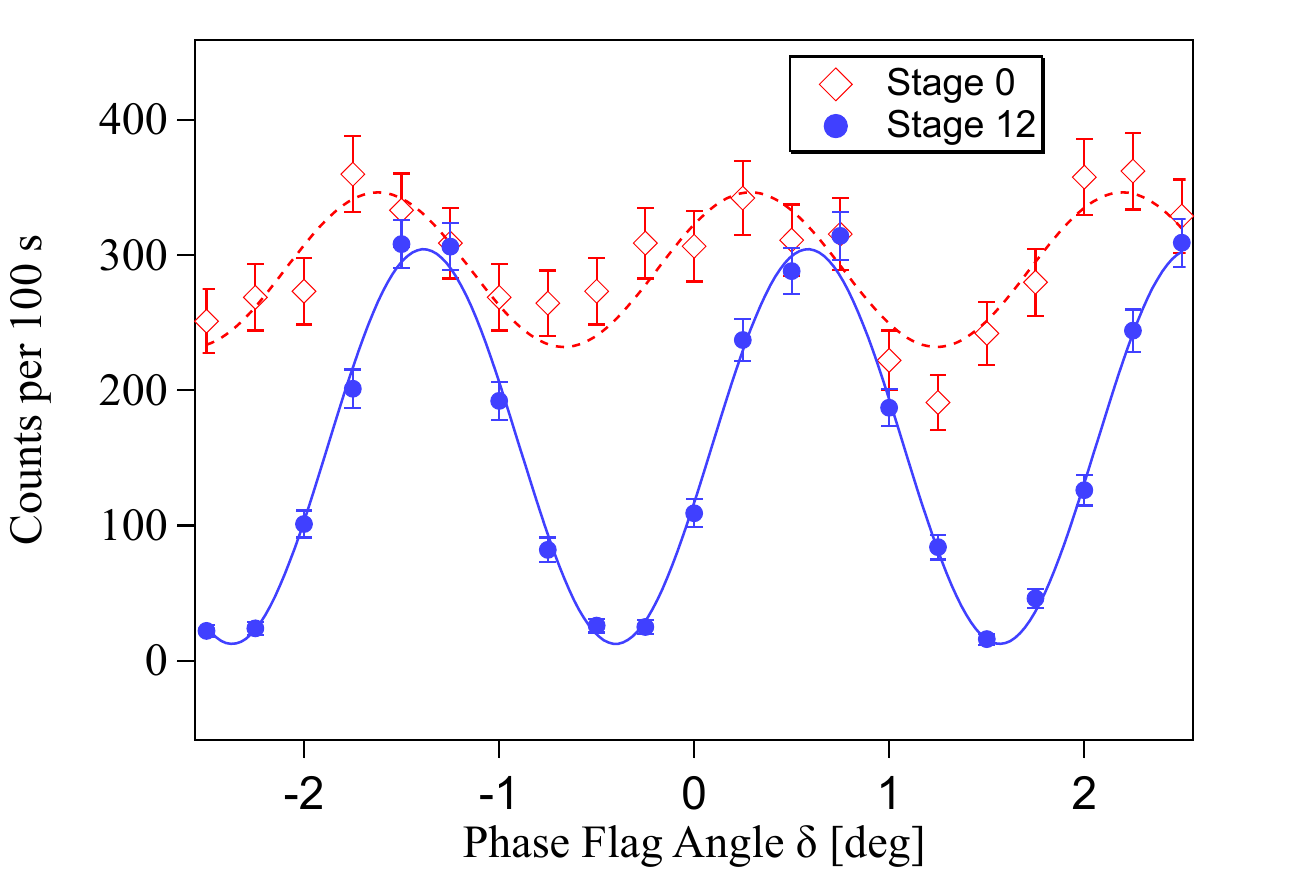} 
 \caption{Interferograms of the maximum contrast, or fringe visibility, observed in the $I_O$ beam for the initial and final  stage (see Table \ref{table1}) of fabrication.  The \textcolor{red}{$\diamond$} (~\filledcirc~) symbols represent the $I_O$ intensity for the initial (final) stage.  Lines are fits to the data using Eqn.~\ref{fit1} and the uncertainties shown are purely statistical.  See the text for a description the techniques used at each fabrication stage.  }
 \label{infero}
\end{figure}

 \begin{figure}[t]
 \centering
\includegraphics[trim={100 2400 1200 10},clip,width = \columnwidth]{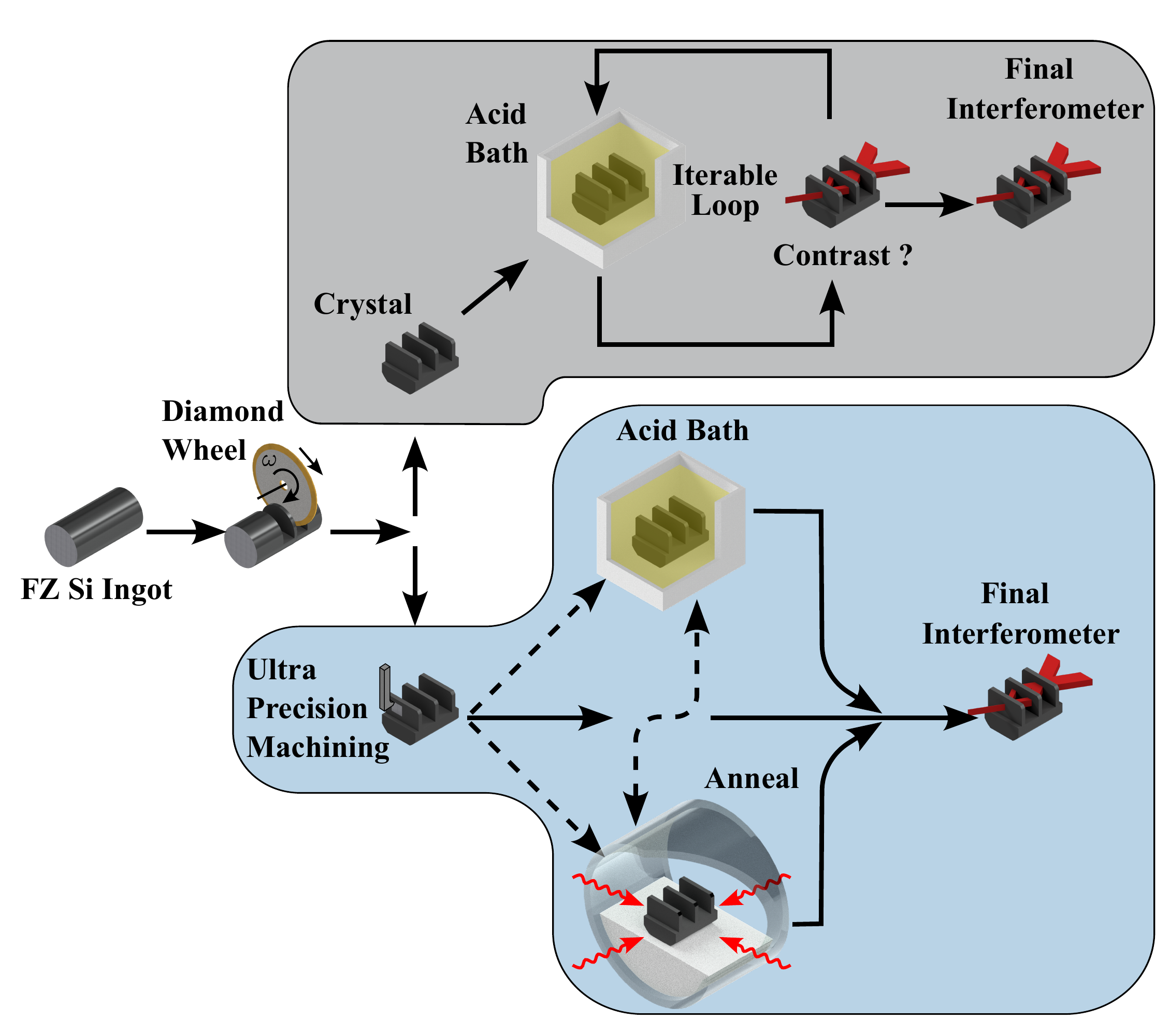} 
 \caption{The fabrication processes.  The common pathway to fabricate an interferometer crystal is highlighted in grey.  Here a float zone silicon ingot is machined and then etched.  The etching process is repeated until the measured contrast of the interferometer no longer improves with each etch.  In our process (blue) after a rough cut of the ingot, more precise and less damaging machining is done.  Optional steps are depicted by the dotted arrows. Note that no more than one etch is required.}
 \label{fabprocess}
\end{figure}

The intensity measured by the neutron detectors is a function of the phase difference between the two paths, $\Delta\phi = \phi_\mathrm{II} -\phi_\mathrm{I}$, and is given by
\begin{eqnarray}
    I_O \propto  \abs{\Psi_\mathrm{I}e^{i\phi_\mathrm{I}} + \Psi_\mathrm{II}e^{i\phi_\mathrm{II}}}^2 \propto [1 +\cos(\Delta\phi)].
\end{eqnarray}
In practice, $I_O$ and $I_H$ are fit to the functions
\begin{eqnarray}
I_O &=& c_0 + c_1 \cos[\Delta\phi+\Delta\phi_{0}+\Delta\phi_\mathrm{pf}(\delta)] \label{fit1}  \\
I_H &=& c_2 - c_1\cos[\Delta\phi+\Delta\phi_{0}+\Delta\phi_\mathrm{pf}(\delta)]. \label{fit}
\end{eqnarray}
where $c_0$, $c_1$, $c_2$, and $\Delta\phi_{0}$ are fit parameters.
Note that $I_H + I_O = c_0+c_2 = $ constant; meaning that no neutrons are created or destroyed in the interferometer.  
The phase $\Delta\phi_{0}$ is the phase shift of an empty interferometer and
can vary over time due to external factors like ambient temperature fluctuations \cite{Saggu_2016_Rev}. 
The phase due to the phase flag $\Delta\phi_\mathrm{pf}(\delta)$ is varied by rotating the flag by an angle of $\delta$.
The phase $\Delta\phi$ is the parameter of experimental interest and can be due to a physical sample inside the interferometer or a field depending on the configuration of the apparatus. 
The intensity as a function of $\delta$ is called an interferogram.  
An interferogram showing the initial and final results of this work is shown in Fig.~\ref{infero}.

The contrast, or fringe visibility, is defined for the $I_{O,H}$ beams as
\begin{eqnarray}
\mathcal{C} &=& \frac{\text {amplitude}}{\text {mean}} = \frac{c_1}{c_{0,2}}. \label{con}
\end{eqnarray}
The contrast is a useful quantity for determining the quality of an interferometer.
Ideally,  $\mathcal{C} = $ \unit[100]{\%} in the O-beam.
However, crystal imperfections, internal stresses, surface roughness, and geometric inaccuracies cause the contrast of practical interferometers to be less than \unit[100]{\%} \cite{Nsofini_2016_Phys,Nsofini_2017_J}.
The reflection and transmission coefficients for diffracting into the H-beam limits the maximum H-beam contrast to $\approx $ \unit[40]{\%} \cite{Rauch2015}.

\subsection{Test Interferometer: Talos}
The interferometer called Talos was created over 20 years ago using scrap FZ material. 
It is a LLL type interferometer with a (111) Bragg vector parallel to the blade surface. 
Physically, Talos is closer in size to a perfect-crystal x-ray interferometer \cite{Deslattes1974} than that of a working neutron interferometer.
However, it is of the same geometry and Bragg vector as our most utilized neutron interferometer.  
Talos has a blade-to-blade separation of $L = $ \unit[10]{mm} with a blade height of \unit[15]{mm}.
The blades began this work $D \approx $ \unit[1]{mm} thick.
Talos successfully demonstrated contrast after what was described as a ``quick-and-dirty" fabrication process.  
Its initial contrast was around \unit[30]{\%} but it has fallen lower throughout the years (a phenomenon sometimes observed with other crystals as well).
\par
Due to its relatively small size and thus the limited sample size that it can accommodate, Talos has rarely been used at the facility.  
One exception, where its size was an advantage, was in an attempted effort to buoyantly float an interferometer crystal \cite{Kaiser2006}. 
Neutrons were an early test-bed of the gravitationally induced quantum interference and the equivalence principle \cite{Staudenmann1980,Werner1988,Arif1994,Littrell1997} in a series of so-called ``COW" experiments, a name derived from the original author's last names \cite{Colella1975}.  
Each measurement of gravity using neutrons was systematically limited and disagreed with the theoretically expected result by several percent. However, each new neutron measurement improved on systematics. 
The last COW measurement determined $\Delta\phi_{g,\mathrm{expt}} = $ \werb{210.28}{0.23}{rad$/$\AA} by rotating an interferometer crystal to an angle of \unit[$\pi/4$]{rad}.
This result still disagreed with the predicted value of $\Delta\phi_{g,\mathrm{theo}} = $ \werb{212.119}{0.021}{rad$/$\AA} by \unit[0.8]{\%}, but it had cut the discrepancy by a factor of 2.5 from the previous COW result.
Suspending the crystal in a buoyant liquid was attempted in order to completely eliminate systematics caused by the induced stresses as the crystal is rotated. 
The size of Talos allowed testing of this suspension using a small volume of liquid. 
Ultimately this approach was not pursued and currently the equivalence principle has been established at the $10^{-12}$ level through atom interferometry \cite{Asenbaum2020}.
\par
For the last several years, Talos has been used to test various fabrication techniques.  
The advantages of Talos are that it had established non-zero contrast, its small area allows for faster turnaround, and since it has struggled to find uses elsewhere, permanent destruction of it would not be a major disruption to activities at the facility.
Over 12 fabrication stages Talos' surfaces were either re-fabricated, the crystal etched, or the crystal was annealed. 
The resulting fabrication process is shown in Fig.~\ref{fabprocess}.
At each fabrication stage the performance of Talos was measured using neutrons.

\section{Experiment}
\subsection{Facilities}
\begin{figure}[t]
 \centering
\includegraphics[width = \columnwidth]{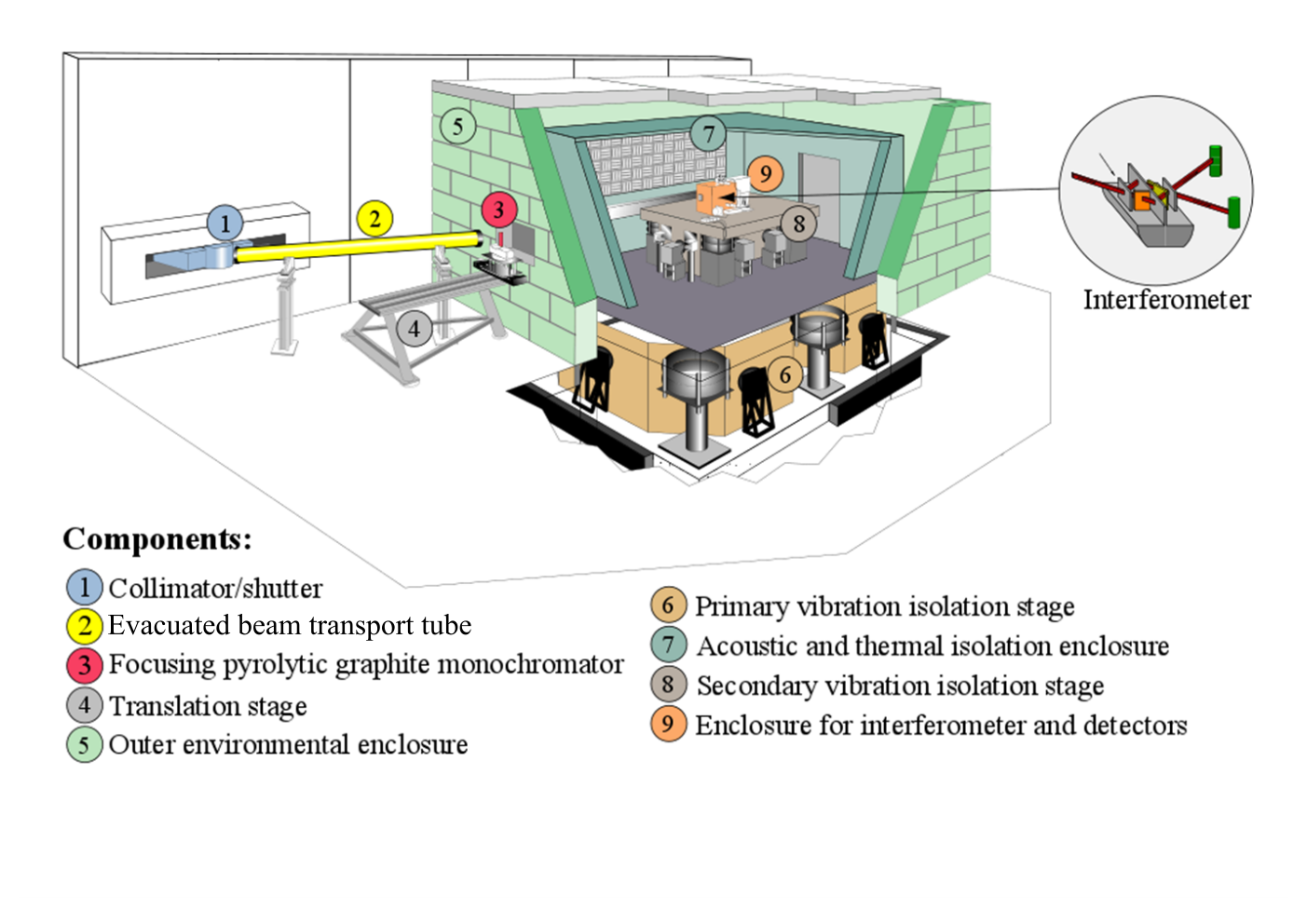} 
 \caption{The Hutch at the \NIOF~provides an isolated environment for stable neutron interferometry measurements. }
 \label{hutch}
\end{figure}

The neutron measurements and annealing were carried out at the \NCNR~in Gaithersburg, MD. 
The \NCNR~provides neutrons for scientific research using a heavy water moderated 20 MW reactor.
There are two operating \NIOF~at the \NCNR~\cite{Shahi2016,Pushin2014}. 
The contrast of Talos was measured inside the Hutch (see Fig. \ref{hutch}) where vibration and thermal isolation allow for achieving maximum contrast under the best possible experimental conditions.
The neutron wavelength used here was $\lambda =$ \unit[2.71]{\AA} with a corresponding Bragg angle of \unit[0.447]{rad} (\unit[25.6]{\degree}). 
The incident beam was \unit[2]{mm} in height and either \unit[2]{mm} or \unit[1]{mm} wide.
Generally, a smaller incident beam will have a larger contrast.
The interferometer was supported by a rotation stage, a y-translation stage, and z-stage.
Because Talos' blade-to-blade distance $L$ is so small,  the forward diffracted beams from the mirror crystal ($I_{1}$ and $I_{4}$) can be reflected in the analyzer crystal and reach the detectors.  
To prevent this, neutron absorbing cadmium was placed behind the analyzer crystal.
\par
The set of high precision fabrication and several etching steps were performed by Neutron Optics INC. \cite{NIST_disclamer}. An ultra precision machining center \cite{VV}, typically used for single point diamond turning optical elements, was modified and adapted for a variety of fabrication methods. 
\par
\begin{figure}[t]
 \centering
\includegraphics[trim={0 1140 0 0},clip,width = \columnwidth]{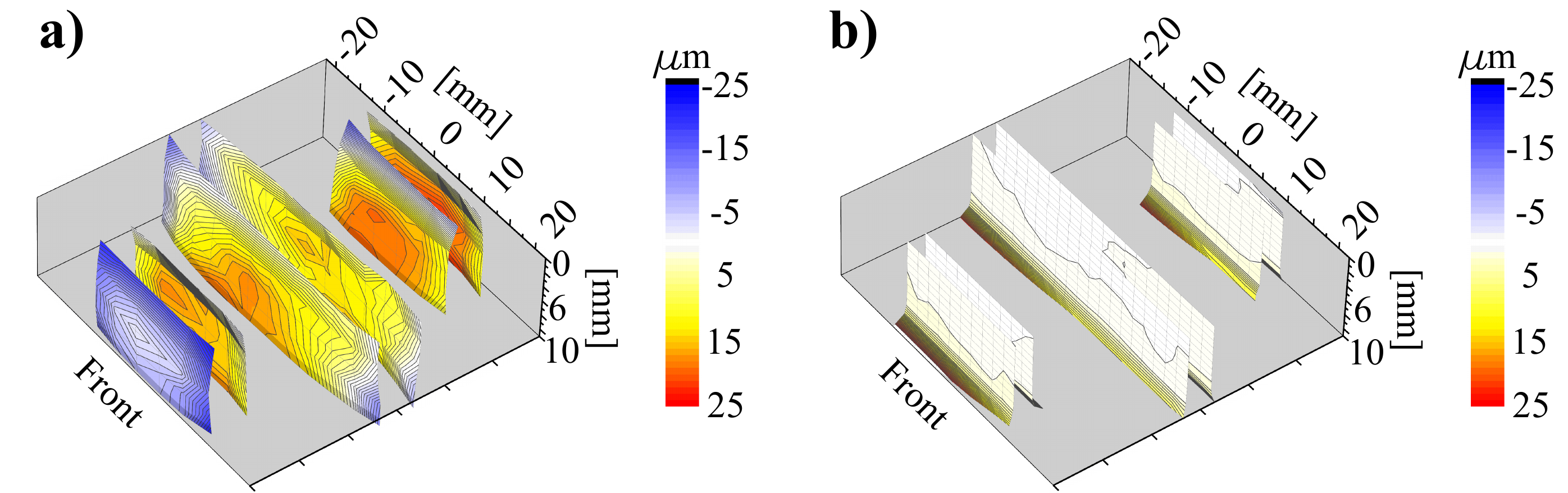} 
 \caption{Front and back blade dimensional measurements of Talos taken at (a) fabrication stage 0 and (b) fabrication stage 10 (see Table \ref{table1}).  Values are relative to an ideal geometry with $L =$ \unit[10.855]{mm} and  $D =$ \unit[0.870]{mm}. Measurements were made at \unit[21]{$^\circ$C}. Contour lines are spaced by \unit[1]{{\textmu}m}.
}
 \label{profiles}
\end{figure}

\begin{figure}[t]
 \centering
\includegraphics[trim={0 330 0 0},clip,width = \columnwidth]{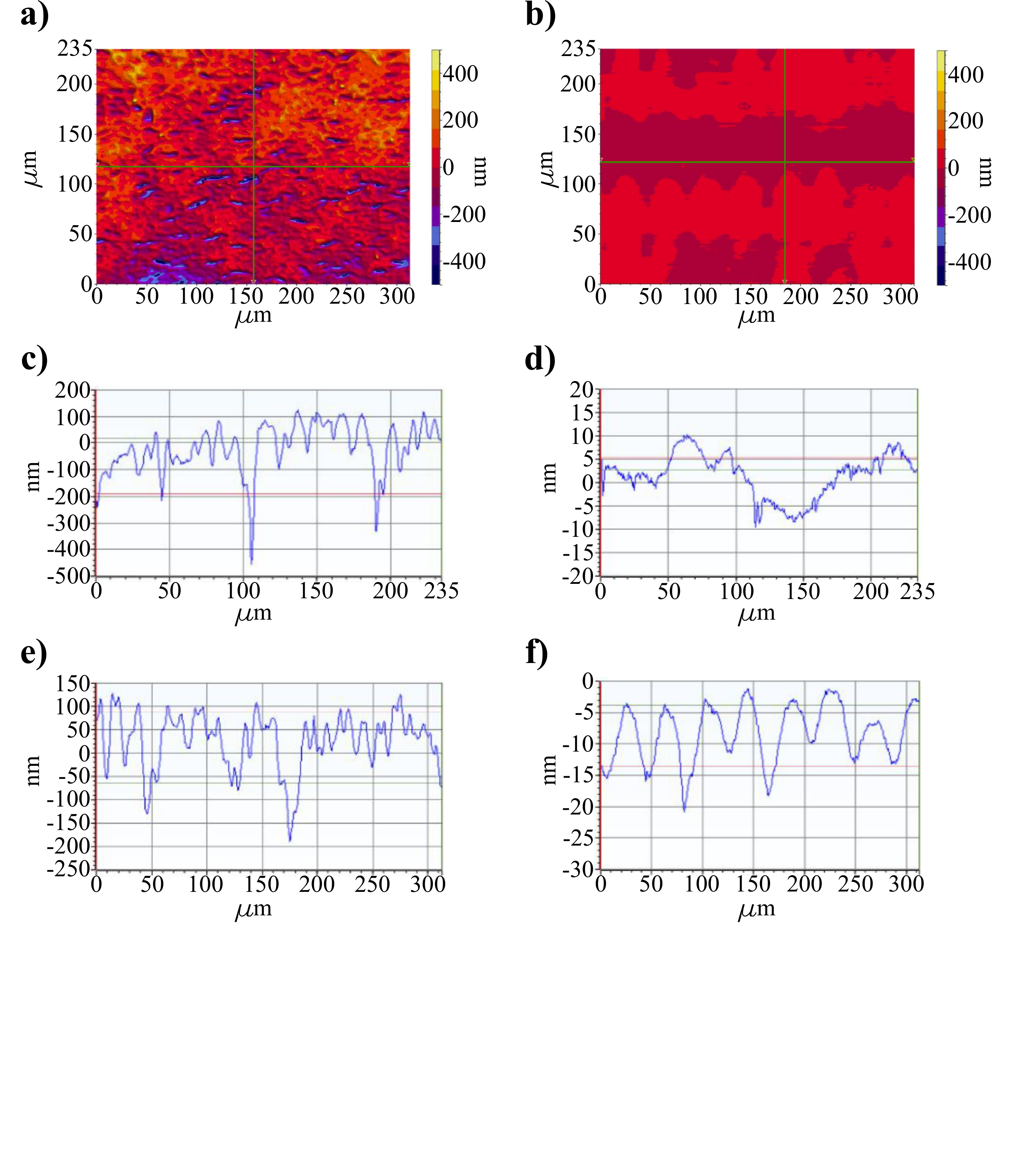} 
 \caption{ (a-b) Measurements showing local variation of the surface height of Talos. The left column was taken at fabrication stage 0 and the right column at stage 10 (see Table \ref{table1}).  (c-f) Line profiles from (a) and (b) showing the variation.  Measurements were made at \unit[21]{$^\circ$C}.
}
 \label{Lineprofiles}
\end{figure}

 
The first step was to bring the initial geometric tolerance (Stage 0, Table \ref{table1}) of Talos (Fig.~\ref{profiles}(a)) over an order of magnitude closer to an ideal geometry (equal blade thickness $D$ and separation $L$), each subsequent method strived to have a tolerance similar to the one from the stage seen in Fig.~\ref{profiles}(b). Precision measurements of the interferometer geometry were performed via a coordinate measuring machine (CMM)-style contact measurement using an amplified strain gauge in situ and without having to remove the crystal or alter the setup for cutting. The tip is a modified \unit[1]{mm} ruby sphere with a dedicated smoothed flat area free from abrupt edges preventing Hertzian contact. In order to minimize risk of damage, the contact force of \unit[2]{g} never exceeds the yield point of silicon. The measurements were checked for temperature stability and could be compensated in case of linear drift when a time-consuming, dense grid of measurements is desired. A precision silicon nitride reference ball is used for absolute measurements and calibration. The gauge is repeatable to \unit[0.05]{\micro m}. Typical surface measurements on Talos were determined to be accurate to $\pm$ \unit[0.2]{\micro m} at \unit[21]{$^\circ$C}, consist of about 50 total points and took approximately 10 minutes per surface. The fabrication stages, categorized into resurfacing (R), etching (E) and annealing (A), are summarized in Table \ref{table1}.
Figure \ref{Lineprofiles} shows the variation in surface height deceased by a order of magnitude following the ultra precision machining.
\par
The method in stage 1 consisted of a lapping method using a self-charging brass lap submerged in a \unit[0.5]{\micro m} particle size diamond slurry. Fabrication stage 3 utilized a polishing technique resembling a chemo-mechanical polishing process \cite{WW}. Stage 6 employed a water-cooled ‘pinch ruling’ process \cite{XX}, where each blade is cut by two opposing single point diamond cutters in the ductile regime \cite{YY}. Stages 9 and 10 were experimental methods based on elastic emissions machining \cite{ZZ} showing considerable contrast without the need for etching or annealing. 
\par
Post fabrication annealing took place in a large commercial tube furnace.
The interferometer was placed on a graphite crucible centering the crystal in the oven bore.
The oven's temperature was raised at a rate of \unit[1$^\circ$]{C/min} to \unit[200]{$^\circ$C} and held at that temperature for \unit[2]{h}.  
It was then raised to a final annealing temperature of \unit[800]{$^\circ$C} at a slower rate of \unit[0.5]{$^\circ$C/min}.
The crystal was held at the final temperature for \unit[12]{h} and then lowered back to room temperature at a rate of \unit[-0.5]{$^\circ$C/min}.
Annealing has improved the contrast of several interferometers \cite{Heacock2018}, most notably Hera. 
However, other interferometers, namely Dionysus and Achilles, have  had no noticeable change in contrast post annealing.  
For the first time,  annealing has been shown to have negative effects on the interferometer overall performance in stages 4 and 9.  

\subsection{Contrast \& Phase Maps}
\begin{figure}[t]
 \centering
\includegraphics[width = \columnwidth]{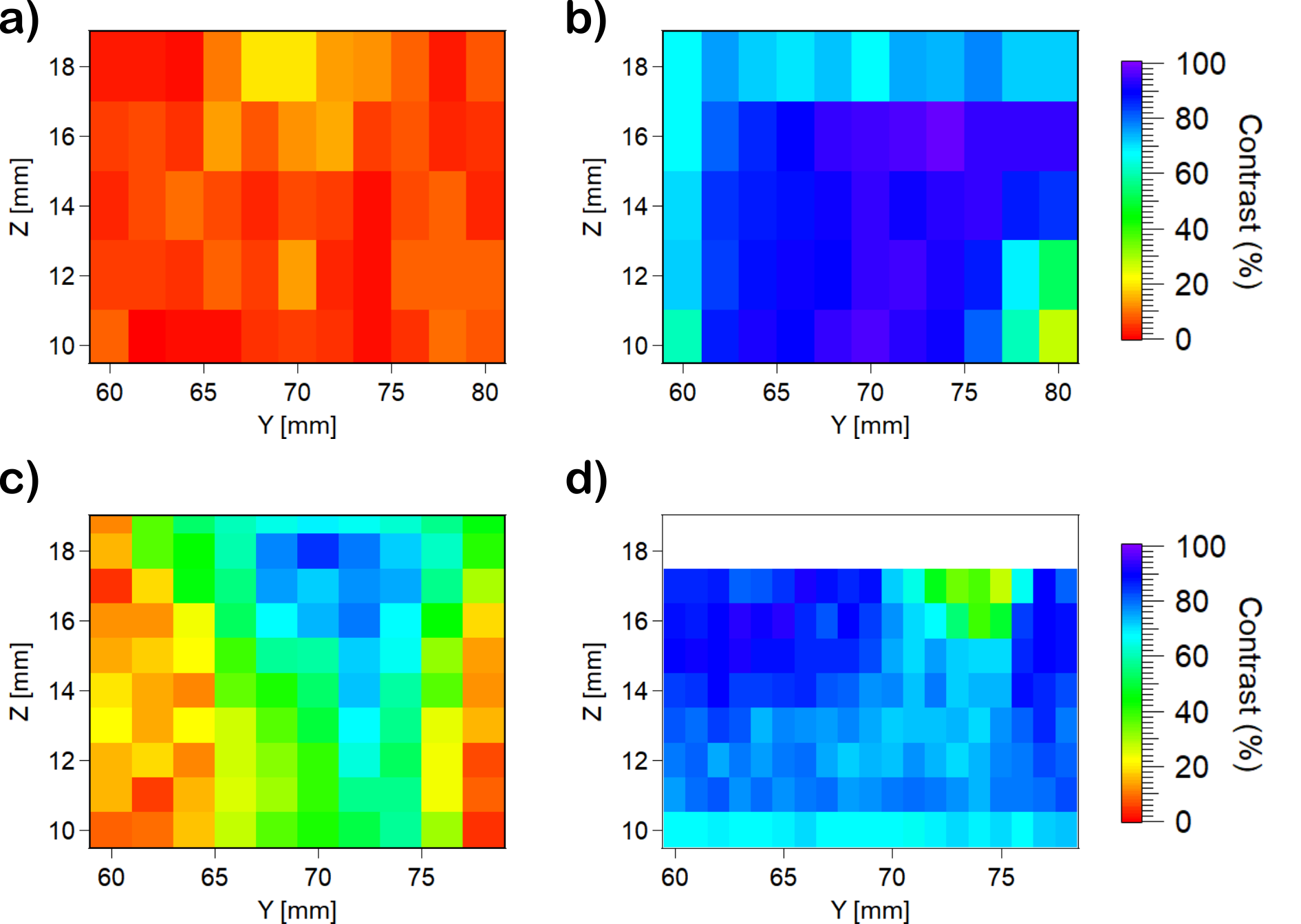} 
 \caption{Contrast maps of Talos. The color scale is the same throughout.  (a) Stage 0; an early contrast map of the interferometer. (b) Stage 2; having no measurable contrast followed by a 5 minute acid etch.  (c) Stage 4; after the results in (b) the faces of the interferometer were re-fabricated and the crystal was annealed.  Contrast decreased and a sweet spot appeared. (d) Stage 12; high contrast as restored after re-fabricating the crystal surfaces.  No data was taken at coordinate Z $=$ \unit[18]{mm}. }
 \label{BabyContrast}
\end{figure}

To measure interferometer contrast the crystal is translated vertically and horizontally allowing the incident beam to illuminate different points along the crystal face. 
At each point the phase flag is rotated $\pm$ \unit[2.5]{\degree} creating an interferogram like those in Fig.~\ref{infero} from which the contrast can be determined via Eqn.~\ref{fit1}.
Contrast maps of four stages are shown in Fig.~\ref{BabyContrast}.
An interferometer can  exhibit a strongly preferred y-z position called a sweet spot.
An example of this can be seen in Fig.~\ref{BabyContrast}(c).
\par
\begingroup

\setlength{\tabcolsep}{6pt} 
\renewcommand{\arraystretch}{1.1} 
\begin{table}[!b]{\caption{Maximum Contrast and the contrast variation of Talos after various processes. Note Stage 4 and 5 were taken using a narrower slit (\unit[2]{mm} $\times$ \unit[1]{mm}) which likely increased the contrast measured during those runs.   Key: Resurfaced (R), Annealed (A), Etched (E) }\label{table1} }
\centering  
\begin{tabular}[c]{lccr}
	\hline\hline
        Fab. &Process &    \multicolumn{2}{c}{Contrast, $\mathcal{C}$}     \\
 Stage  & (See Key) &   Var.  (\%)   &    Max. (\%)  \\
 \hline
        0  & \emph{initial} & 7 $\pm$ 4   & 20 $\pm$ 3 \\
        1  & R \& A            & nil         & nil        \\
        2  & E                 & 82 $\pm$ 13 & 97 $\pm$ 4 \\
        3  & R                 & 61 $\pm$ 21 & 87 $\pm$ 7 \\
        4  & A                 & 33 $\pm$ 20 & 76 $\pm$ 6 \\
        5  & E                 & 5 $\pm$ 27  & 90 $\pm$ 6 \\
        6  & R                 & 4 $\pm$ 5   & 24 $\pm$ 1 \\
        7  & E                 & 22 $\pm$ 15 & 45 $\pm$ 2 \\
        8  & R                 & 9 $\pm$ 10  & 40 $\pm$ 3 \\
        9  & A                 & 3 $\pm$ 2   &  8 $\pm$ 2 \\
        10 & R                 & 34 $\pm$ 16 & 75 $\pm$ 3 \\
        11 & R                 & 7 $\pm$ 5   & 23 $\pm$ 3 \\
        12 & R                 & 71 $\pm$ 18 & 92 $\pm$ 4 \\
	\hline\hline
	\end{tabular}		
\end{table}
\endgroup

Table \ref{table1} shows the results of various processes on the interferometer and its effect on the contrast. 
The maximum contrast observed was over \unit[90]{\%} compared to a minimum of  \unit[8]{\%} following annealing. 
For fabrication stage 2 the crystal was etched in a solution of  
nitric acid to hydrofluoric acid  (HNO$_3$:HF) at a 60:1 volume ratio  for \unit[5]{min} removing approximately \unit[1]{\micro m} of silicon material off each blade.
We were able to destroy and then completely restore contrast of the interferometer several times indicating that the fabrication stages caused no  permanent stress or damage to the crystal.
\par
\begin{figure}[t]
 \centering
\includegraphics[width = \columnwidth]{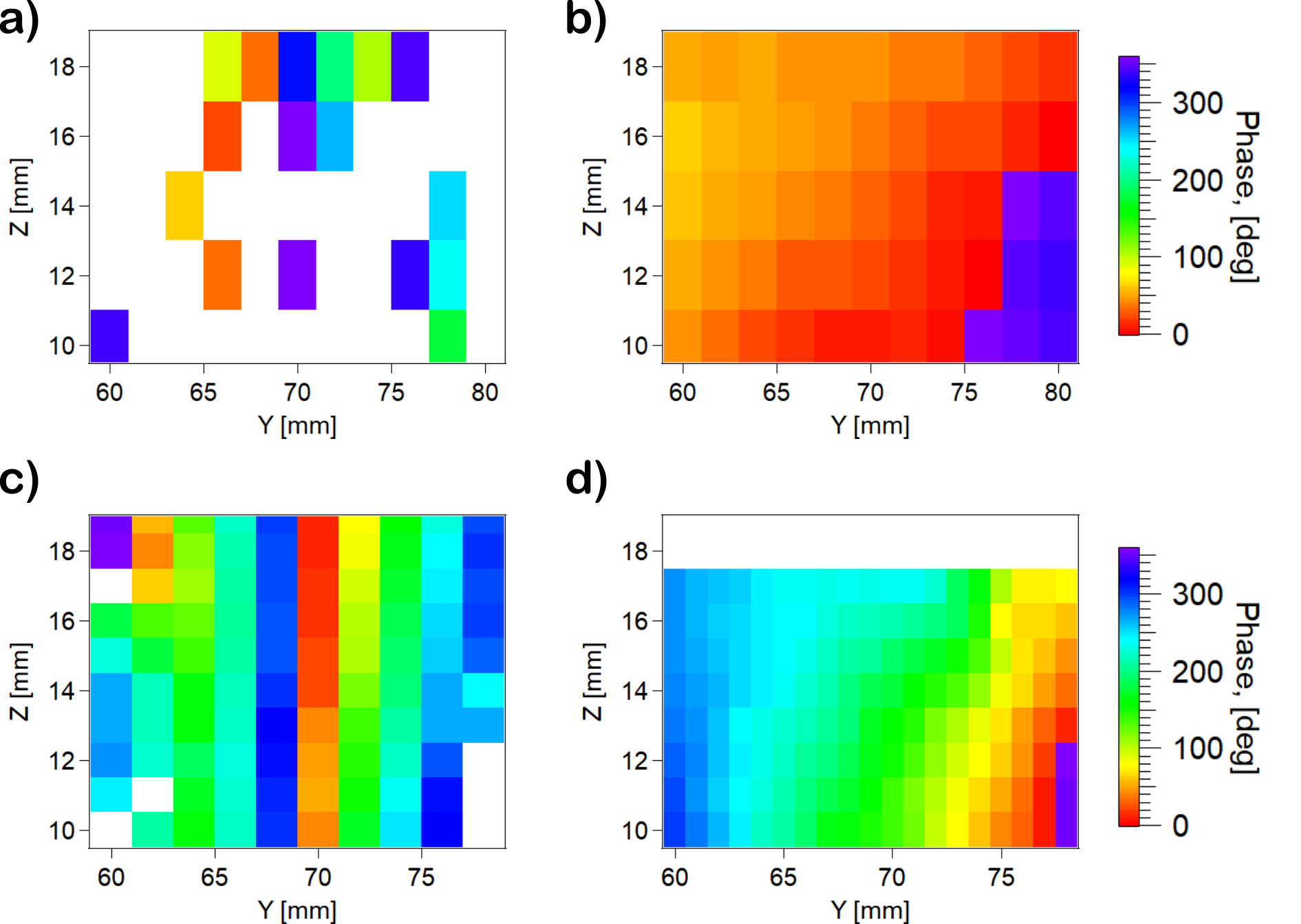} 
 \caption{Phase maps of Talos for the same cases shown in Fig.~\ref{BabyContrast}. The color scale is the same throughout. White pixels in (a) and (c) correspond to areas of low contrast where $\Delta\phi$ uncertainties are greater or equal than \unit[20]{\degree} and have been omitted. No data was taken at coordinate Z $=$ \unit[18]{mm} for (d). In (c) a modulus of $2\pi$ is crossed in the middle.}
 \label{BabyPhase}
\end{figure}

The measured phase $\Delta\phi_{0}$ at each point is plotted in Fig.~\ref{BabyPhase}.
Points of low contrast where the fitted uncertainty $\sigma_{\Delta\phi_{0}} \geq $ \unit[20]{\degree} are omitted (white pixels) in the phase maps. 
Since the interferometer measures $\Delta\phi$ modulo $2\pi$, a $2\pi$ phase step can be seen in the middle of 
Fig.~\ref{BabyPhase}(c).  
A clear, smooth gradient can be seen in the phase maps corresponding to high contrast.  
\par
The contrast variation is defined as the mean contrast $\pm$ 1 standard deviation.
As shown in Fig.~\ref{BabyContrast}(d), Talos doesn't currently exhibit a sweet spot and instead has a roughly uniform contrast map. 
The largest improvements in contrast (stages 2, 5, and 7) followed etching using  HNO$_3$:HF mixture or KOH.
KOH is usually avoided because it etches anisotropically, but was tried here to avoid using more dangerous HF solutions.


\section{Discussion}
In this work we have shown that advanced fabrication using a diamond turning machine produces a high-quality, high-contrast neutron interferometer.
One can achieve this high contrast without the need for performing several lengthy acid etchs. Additionally, for the first time we have shown a negative response to annealing for a neutron interferometer which was improved after an etch. A similar behavior was observed in an x-ray interferometer \cite{Massa_2020_JoAC}.
\par
Neutron interferometry requires highly precise silicon fabrication techniques that can be consistently applied to a relatively large surface area (\unit[12]{cm$^2$}). 
Specifically, the imaging of samples inside an interferometer generally requires using relatively large (\unit[10]{mm} x \unit[10]{mm}) incident beams \cite{Sarenac_2016_Opt.Express,Pushin_2007_Appl.Phys.Lett.}. 
Therefore, sample imaging  requires uniform contrast and phase profiles over large crystal surfaces.  
Our techniques have shown uniformity to a high degree over such areas.  
Higher performance crystals are needed for the next generation of neutron interferometry experiments including a split neutron interferometer \cite{Lemmel2022}, a decoherence-free subspace neutron interferometer\cite{Pushin_2009_PhysRev}, and for future facilities \cite{Jurns_2020}.
\par
In the past, interferometer crystals were cut and then etched in an iterable process until a maximum contrast was observed. 
Here we have eliminated etching and iterated, as a test, the fabrication.
This type of fabrication is much slower, but a much more reproducible and controlled action. 
The reduction or complete elimination of the iterable etching process also removes safety and disposal concerns related to using a large volume of strong acids.
Disposal concerns are especially consequential for fabricating an interferometer made of germanium; a material that becomes radioactive during every iterable contrast map.
In the near future, a large interferometer crystal, silmular in geometry to Hera (see Fig.~\ref{inters}), using these new techniques will be deployed at the NIST interferometer facilities.

\section{Acknowledgements}

This work was supported by the Canadian Excellence Research Chairs (CERC) program, the Natural Sciences and Engineering Council of Canada (NSERC) Discovery program, the Canada First Research Excellence Fund (CFREF), Triangle Universities Nuclear Laboratory (TUNL), the National Institute of Standards and Technology (NIST) and the US Department of Energy, Office of Nuclear Physics, under Interagency Agreement 89243019SSC000025. 

\bibliographystyle{elsarticle-num}  
\bibliography{TalosFab.bib}

\begin{thebibliography}{10}
\bibitem{NIST_disclamer}
Certain trade names and company products are mentioned in the text or
  identified in an illustration in order to adequately specify the experimental
  procedure and equipment used. In no case does such identification imply
  recommendation or endorsement by the National Institute of Standards and
  Technology, nor does it imply that the products are necessarily the best
  available for the purpose.
\end{thebibliography}

\end{document}